\documentclass[conference]{IEEEtran}
\IEEEoverridecommandlockouts
\usepackage{shortcuts_OPT}
\usepackage{cite}
\usepackage{amsmath,amssymb,amsfonts}
\usepackage{algorithmic}
\usepackage{graphicx}
\usepackage{epstopdf}
\usepackage{textcomp}
\usepackage{color}
\usepackage{subfigure}
\def\revcolor{\color{black}}
\def\BibTeX{{\rm B\kern-.05em{\sc i\kern-.025em b}\kern-.08em
    T\kern-.1667em\lower.7ex\hbox{E}\kern-.125emX}}
    
\begin{document}

\title{Multilayer Simplex-Structured Matrix Factorization for Hyperspectral Unmixing with Endmember Variability
\thanks{This work was supported by a General Research Fund (GRF) of Hong
Kong Research Grant Council (RGC) under Project ID CUHK 14203721.}
}

\author{\IEEEauthorblockN{Junbin Liu, Yuening Li, and Wing-Kin Ma}
\IEEEauthorblockA{\textit{Department of Electronic Engineering, The Chinese University of Hong Kong, Hong Kong SAR of China}\\
liujunbin@link.cuhk.edu.hk, yuening@link.cuhk.edu.hk, wkma@ee.cuhk.edu.hk}
}

\maketitle

\begin{abstract}
Given a hyperspectral image, the problem of hyperspectral unmixing (HU)
is to identify the endmembers (or materials) and the abundance (or
endmembers' contributions on pixels) that underlie the image. 
HU can be
seen as a matrix factorization problem with a simplex structure {\revcolor with} the
abundance matrix factor. In practice, hyperspectral images may exhibit
endmember variability (EV) effects---the endmember matrix factor varies
from one pixel to another. In this paper we consider a multilayer
simplex-structured matrix factorization model to account for the EV
effects. Our multilayer model is based on the postulate that if we
arrange the varied endmembers as an expanded endmember matrix, that
matrix exhibits a low-rank structure. A variational inference-based
maximum-likelihood estimation method is employed to tackle the
multilayer factorization problem. Simulation results are provided to
demonstrate the performance of our multilayer factorization method.
\end{abstract}


\section{Introduction}
A hyperspectral image is a collection of electromagnetic spectrums within the surveillance range.
Because of the spatial resolution limitation, one hyperspectral image pixel can have several materials' electromagnetic reflectances, termed spectral signatures or endmembers (EMs), mixed together.
Hyperspectral unmixing (HU) is an important development in hyperspectral imaging aiming at identifying the mixed materials and their proportions.
HU has been studied for decades, and it is typically based on the so-called {\revcolor linear} mixture model (LMM) \cite{keshava2002spectral}.

An underlying assumption of the LMM is that each material is perfectly represented by a single endmember. 
This assumption is strong as the endmembers corresponding to the same material can vary across pixels due to, e.g., the change of illumination and atmospheric conditions \cite{healey1999models,zare2013endmember,borsoi2021spectral}. 
Such an endmember-varying phenomenon is called endmember variability (EV).
Ignoring EV can lead to performance loss.
This motivates efforts to develop algorithms enabling EMs to vary within a hyperspectral image.
There are various methods based on different EM models.
For example, the extended LMM models capture the scaling variability of EMs, while probabilistic methods model EM variants as samples from some distributions. 
There are also some methods, for instance, random search methods \cite{roberts1998mapping} and sparse HU methods \cite{bioucas2010alternating,iordache2011sparse,xu2018l0} that rely on an external EM library.

In this paper, we propose a different strategy to model the EV.
We concatenate the various EMs into an expanded EM matrix, and we assume the matrix exhibits a low-rank structure.
This is motivated by the fact that variants of EMs 
from the same material are similar.
While there can be other possibilities to model low-rank structures,
we perform the decomposition of the low-rank expanded EM matrix and develop a multilayer simplex-structured matrix factorization model.
Maximum-likelihood estimation is used to infer the EMs with variability under the multilayer model.
Experiments on a real hyperspectral image reveal interesting results about how the EMs {\revcolor may be} hierarchically represented.

We should note that, recently, there has been interest in using multilayer structured matrix factorizations to model low-rank structures and uncover hierarchical and structural data information, see e.g., \cite{trigeorgis2016deep, de2021survey}.
Multilayer matrix factorization models have been applied to various applications such as classification \cite{song2015hierarchical}, clustering \cite{de2021deep} and multi-view clustering \cite{wei2020multi}, and HU without EV \cite{rajabi2014spectral}.
To the best of our knowledge, no multilayer matrix factorization model has been developed for HU {\revcolor to} account for EV.

\section{Background}

We start with the basic hyperspectral unmixing (HU) problem.
We obtain a hyperspectral measurement of a scene or a hyperspectral image.
The hyperspectral image is modeled by a linear mixture model (LMM)
\begin{equation}\label{eq:LMM}
\boldsymbol{y}_n = \boldsymbol{As}_n +\boldsymbol{v}_n, \quad n=1, \ldots, N,
\end{equation} 
where $\boldsymbol{y}_n \in \mathbb{R}^M$ denotes the $n$th pixel of the hyperspectral image;
$M$ denotes the number of spectral bands of the hyperspectral image;
$\boldsymbol{A} \in \mathbb{R}^{M \times K}$ is an endmember (EM) matrix in which each column $\boldsymbol{a}_i$ of $\boldsymbol{A}$ represents the spectral signature, or EM, of a distinct material;
$K$ denotes the total number of EMs (or materials) in the image;
$\boldsymbol{s}_n$ denotes the abundance of the various EMs at the $n$th pixel;
$\boldsymbol{v}_n$ represents noise.
In particular, the LMM has the physical interpretation that we see each pixel of the hyperspectral image as a linear combination of various EMs (or materials).
It is common to assume that each pixel is a proportional, or convex, combination of the EMs; that is, every $\boldsymbol{s}_n$ is non-negative and has the sum-to-one property $\boldsymbol{s}_n^\top \boldsymbol{1} = 1$ (\ie $\boldsymbol{s}_n$ lies in the unit simplex).
The reader is referred to the literature, such as \cite{Ma2014HU,Jose12}, for modeling details.
The problem of HU is to identify the EM matrix $\boldsymbol{A}$ and the abundances $\boldsymbol{s}_n$'s from the hyperspectral image $\boldsymbol{Y}= [~ \boldsymbol{y}_1,\ldots,\boldsymbol{y}_N ~]$.
From the problem statement, we may see HU as a simplex-structured matrix factorization problem.
In fact, some HU algorithms are built from the following formulation
\beq \label{eq:vol_reg}
\min_{\boldsymbol{A}, \boldsymbol{S}} \| \boldsymbol{Y} - \boldsymbol{A} \boldsymbol{S} \|_F^2 + r(\boldsymbol{A})
~~ {\rm s.t.} ~~ \boldsymbol{S} \geq \boldsymbol{0}, \boldsymbol{S}^\top \boldsymbol{1} = \boldsymbol{1},
\eeq 
where $r$ denotes a regularizer, a common one being the volume regularizer $r(\boldsymbol{A}) = \lambda \log( \det(\boldsymbol{A}^\top \boldsymbol{A}) )$ for some constant $\lambda > 0$; see, e.g., \cite{miao2007endmember,fu2016robust}.

The LMM is simple and widely used.
It however does not take the endmember variability (EV) effects into consideration.
In reality, EMs can have pixel-wise variations due to illumination effects, atmospheric conditions, and materials' intrinsic structural differences \cite{healey1999models,zare2013endmember,borsoi2021spectral}.
Neglecting the EV effects can lead to performance {\revcolor loss}.
Under EV, the LMM should be modified as
\beq
\boldsymbol{y}_n = \boldsymbol{A}_n \boldsymbol{s}_n+\boldsymbol{v}_n, \quad n=1, \ldots, N,
\eeq 
where $\boldsymbol{A}_n \in \mathbb{R}^{M \times K}$ describes an EM matrix that can vary across pixels.
There are a wide variety of models for $\boldsymbol{A}_n$; see, e.g., \cite{nascimento2005does,drumetz2016blind,drumetz2020spectral,hong2018augmented,stein2003application,du2014spatial,li2020stochastic}.
For instance, according to the aforementioned literature, we may model $\boldsymbol{A}_n$ as
\beq
\boldsymbol{A}_n = \boldsymbol{A} \boldsymbol{C}_n + \boldsymbol{E}_n,
\eeq 
where $\boldsymbol{A} \in \mathbb{R}^{M \times K}$ describes the base EM matrix;
$\boldsymbol{C} \in \mathbb{R}^{K \times K}$ is a positive diagonal matrix that models scaling effects with the EMs, or the so-called ``extrinsic'' EM effects;
$\boldsymbol{E}_n \in \mathbb{R}^{M \times K}$ represents variability that cannot be represented by scaling, which is called the ``intrinsic'' EM effects.
As an alternative to the EV model in (4), one may use a probabilistic model to model $\boldsymbol{A}_n$\cite{stein2003application,du2014spatial,li2020stochastic}.
The spirit of such probabilistic models is similar to that in (4) in the sense that we model $\boldsymbol{A}_n$ as some form of variation from the base EM matrix.
The matrix factorization problem associated with the LMM under EV is more complex, requiring us to take the model of $\boldsymbol{A}_n$ into account; see \cite{nascimento2005does,drumetz2016blind,drumetz2020spectral} and also \cite{stein2003application,du2014spatial,li2020stochastic} for non-matrix factorization methods.

The aforementioned EV-handling approach requires us to put down a model for the pixel-variant EM matrix $\boldsymbol{A}_n$.
There is a different approach that does not use a model.
The approach assumes that we have an external EM library, denoted by $\boldsymbol{A}_{\rm lib} \in \mathbb{R}^{M \times K_{\rm lib}}$.
Here, each column of $\boldsymbol{A}_{\rm lib}$ is the spectral signature of a material.
Also, $\boldsymbol{A}_{\rm lib}$ contains numerous materials' EMs---including various variations of EMs that come from one material.
The idea is to model the hyperspectral image as 
\beq
\boldsymbol{y}_n = \boldsymbol{A}_{\rm lib} \boldsymbol{s}_n + \boldsymbol{v}_n, \quad n=1,\ldots,N,
\eeq 
where $\boldsymbol{A}_{\rm lib}$ serves as an expanded EM matrix for the hyperspectral image;
$\boldsymbol{s}_n \in \mathbb{R}^{K_{\rm lib}}$ is an expanded abundance and is assumed to be sparse.
The above model means that each hyperspectral image is a combination of a small number of EMs from the EM library.
The problem then becomes a sparse regression problem; 
see \cite{bioucas2010alternating,iordache2011sparse,xu2018l0} for details.

\section{Multilayer Model for HU with EV}

We consider a multilayer model to deal with EV.
The modeling idea is simple, but, to the best of our knowledge, was not previously considered in the context of HU.
We use the same LMM in (\ref{eq:LMM}), \ie
\begin{equation}\label{eq:LMM_lowrank}
\boldsymbol{y}_n = \boldsymbol{As}_n +\boldsymbol{v}_n, \quad n=1, \ldots, N,
\end{equation}   
but now $\boldsymbol{A}$ is as an {\em expanded} EM matrix---which encompasses all the variations of EMs from the base EMs (or base materials).
This resembles the EM library-based model in (5), but we assume that we do not have an EM library. 
Instead, we want to identify $\boldsymbol{A}$ from the hyperspectral image $\boldsymbol{Y}$.
At first, the problem seems to be the same as the standard HU problem (or HU without EV): identify $\boldsymbol{A}$ and the $\boldsymbol{s}_n$'s from $\boldsymbol{Y}$.
But there is a caveat.
The expanded EM matrix $\boldsymbol{A}$ is likely to be either rank-deficient or ill-conditioned because $\boldsymbol{A}$ contains highly similar EMs as variations of the base materials.
In HU or in simplex-structured matrix factorization, 
we typically want $\boldsymbol{A}$ to have full-column rank,
and we may be faced with performance deterioration as $\boldsymbol{A}$ becomes more ill-conditioned.
This means that direct application of HU or simplex-structured matrix factorization methods to the expanded LMM in (6) may not work.
\begin{figure}[htb]
	\centering
	\subfigure[An example of the expanded matrix.]{
		\begin{minipage}[]{0.22\textwidth}
			\includegraphics[width=1\textwidth]{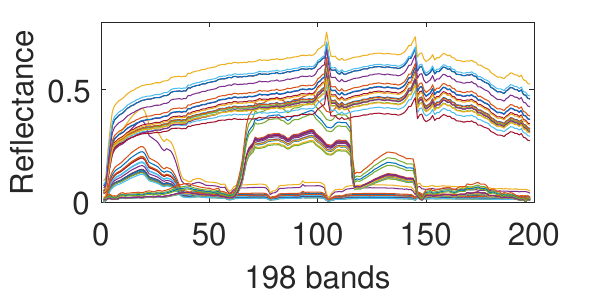}
		\end{minipage}
	}
    	\subfigure[Singular values.]{
    		\begin{minipage}[]{0.22\textwidth}
   		 	\includegraphics[width=1\textwidth]{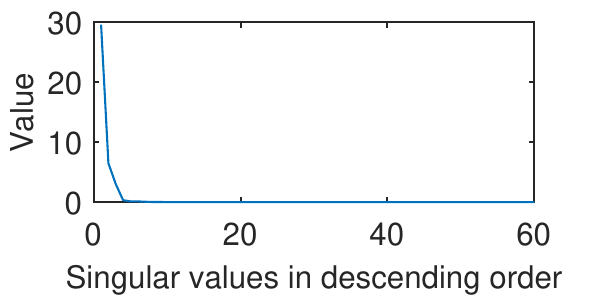}
    		\end{minipage}
    	}
\caption{An example showcases the low-rank structure of an expanded endmember matrix.}
\label{fig:singular_val}
\end{figure}

Our idea is to assume that the expanded EM matrix has low-rank structures.
To describe the idea, consider the illustration in Fig. \ref{fig:singular_val}(a).
We form an expanded EM matrix $\boldsymbol{A}$ by taking different variations of EMs from three base materials;
they were taken from an EM library.
The size of $\boldsymbol{A}$ is $(M,K)= (198,60)$.
Fig. \ref{fig:singular_val}(b) illustrates the singular values of $\boldsymbol{A}$, and we see that $\boldsymbol{A}$ has low-rank structures.
This makes sense since variations of EMs from the same material are highly similar.
Our low-rank model for the expanded EM matrix $\boldsymbol{A}$ is as follows:
\beq\label{eq:factA_2layer}
\boldsymbol{A} = \boldsymbol{A}' \boldsymbol{S}',
\eeq 
where $\boldsymbol{A}' \in \mathbb{R}^{M \times K'}$ and $\boldsymbol{S}' \in \mathbb{R}^{K' \times K}$.
We can impose structures with the latent factors $\boldsymbol{A}'$ and/or $\boldsymbol{S}'$.
In this work we assume that every column $\boldsymbol{s}_n'$ of $\boldsymbol{S}'$ is non-negative and has the sum-to-one property.
This means that every column of the expanded EM matrix $\boldsymbol{A}$ is a convex combination of the columns of $\boldsymbol{A}'$, \ie
$\boldsymbol{a}_n = \boldsymbol{A}' \boldsymbol{s}_n'$, with $\boldsymbol{s}_n'$ lying in the unit simplex.
We can interpret $\boldsymbol{A}'$ as the {\em core basis matrix}---every EM variation $\boldsymbol{a}_n$ of a material comes from a convex combination of the columns of the core basis matrix $\boldsymbol{A}'$.
It is anticipated that some part of the columns of {\revcolor the} core basis matrix $\boldsymbol{A}'$ may contain the base EMs (or ``primal'' spectral signatures of the base materials), while the other part may be perturbation basis vectors that account for the EM variations.
By putting (\ref{eq:factA_2layer}) into (\ref{eq:LMM_lowrank}) and by letting $\boldsymbol{S} = [~ \boldsymbol{s}_1,\ldots,\boldsymbol{s}_N ~]$ and $\boldsymbol{V} = [~ \boldsymbol{v}_1,\ldots,\boldsymbol{v}_n ~]$, we are led to a two-layer simplex-structured matrix factorization (SSMF) model
\beq
\boldsymbol{Y} = \boldsymbol{A}' \boldsymbol{S}' \boldsymbol{S} + \boldsymbol{V},
\eeq 
for which $\boldsymbol{S}$ and $\boldsymbol{S}'$ have their columns lying in the unit simplex.

The idea can be generalized to multiple layers, possibly further capturing intermediate structures and hierarchical EV generation.
Given a number of layers $L$ and a collection of latent factor sizes $N_1,\ldots,N_L$ with $N_L=N$, we consider the following model
\beq \label{eq:mssmf}
\boldsymbol{Y} = \boldsymbol{A}_1 \boldsymbol{S}_1 \cdots \boldsymbol{S}_{L} + \boldsymbol{V},
\eeq
where $\boldsymbol{A}_1 \in \mathbb{R}^{M \times N_1}$ is the core basis matrix;
$\boldsymbol{S}_l \in \mathbb{R}^{N_l \times N_{l+1}}$ has its columns lying in the unit simplex.
By letting $\boldsymbol{A}_l = \boldsymbol{A}_l \boldsymbol{S}_{l-1}$,
we see that the columns of $\boldsymbol{A}_l$ are convex combinations of those of $\boldsymbol{A}_{l-1}$, which forms a hierarchical structure of forming the expanded EM matrix $\boldsymbol{A}_L$ from the core basis matrix $\boldsymbol{A}_1$.
The HU problem with EV under the multilayer SSMF (MSSMF) model is to retrieve $\boldsymbol{A}_1$ and $\boldsymbol{S}_l$'s from $\boldsymbol{Y}$.

\section{Proposed Approach}

Given the multilayer model (\ref{eq:mssmf}), we design an algorithm to retrieve all the factors in this section.
The problem can be formulated as an extension of one-layer matrix factorization problem such as (\ref{eq:vol_reg}), and solve it in an alternating fashion.
Here, we take a probabilistic approach and retrieve the matrix factors via maximum likelihood (ML) estimation.

Specifically, we assume that columns $\boldsymbol{s}_{L,n}$'s of $\boldsymbol{S}_L$ are i.i.d. and uniformly distributed on the unit simplex; {\revcolor the} elements of $\boldsymbol{V}$ follow {\revcolor an i.i.d.} Gaussian distribution with zero-mean and variance $\sigma^2$, \ie
\begin{equation}
    \boldsymbol{s}_{L,n} \sim D(\cdot,\boldsymbol{1}),\quad \boldsymbol{v}_n \sim \mathcal{N}(\boldsymbol{0},\sigma^2\boldsymbol{I}),
\end{equation}
where $\mathcal{D}(\cdot;\boldsymbol{1})$ denotes the uniform Dirichlet distribution.
We denote $\Theta:= \{\boldsymbol{A}_1,\boldsymbol{S}_1,...,\boldsymbol{S}_{L-1}, \sigma^2\}$ as {\revcolor the set of unknown} parameters.
{\revcolor Under the above assumptions, we tackle the MSSMF problem by ML estimation:}
\begin{equation}\label{eq:ML}
\begin{aligned}
      \Theta^\star &= \arg\max_{\Theta\in\mathcal{C}}\frac{1}{N}\sum_{n=1}^N \log p\left(\boldsymbol{y}_n;\Theta\right),
\end{aligned}
\end{equation}
where $\mathcal{C}$ denotes the feasible set of $\Theta$, \ie 
$\mathbb{R}_+^{M\times N_1}$ for $\boldsymbol{A}$, unit simplex for {\revcolor the} columns of $\boldsymbol{S}_l$'s, and $\mathbb{R}_{++}$ for $\sigma^2$;
and $p(\boldsymbol{y}_n;\Theta) = \int p(\boldsymbol{y}_n|\boldsymbol{s}_{L,n};\Theta) p(\boldsymbol{s}_{L,n}) d\boldsymbol{s}_{L,n}$ is the likelihood for each hyperspectral image pixel.
The integration in the likelihood does not possess a closed-form expression, which makes the ML problem (\ref{eq:ML}) difficult to solve directly.

We resort to the variational inference (VI) technique{\revcolor\cite{jordan1999introduction}}. 
VI maximizes the log-likelihood approximately via maximizing a lower bound.
A lower bound of the log-likelihood {\revcolor corresponding} to an image pixel {\revcolor$\boldsymbol{y}$} can be obtained by Jensen's inequality{\revcolor:}
\begin{equation}\label{eq:lb}
\begin{aligned} 
    \log p(\boldsymbol{y};\Theta) 
    &= \log \mathbb{E}_{\boldsymbol{s}\sim q}[p(\boldsymbol{y}|\boldsymbol{s} ;\Theta) p(\boldsymbol{s})/q(\boldsymbol{s} )] \\
    &\geq \mathbb{E}_{\boldsymbol{s}\sim q} [\log p(\boldsymbol{y},\boldsymbol{s} ;\Theta) - \log q(\boldsymbol{s})] ,
\end{aligned}
\end{equation}
where $q(\boldsymbol{s})$ is the so-called variational distribution{\revcolor, which has its support being} the unit simplex.
The equality holds for $q(\boldsymbol{s})=p(\boldsymbol{s}|\boldsymbol{y})$ which, however, cannot be evaluated in closed form.
The variational distribution is set as the Dirichlet distribution parameterized by $\boldsymbol{\beta}_n$, \ie
\begin{equation} q\left(\boldsymbol{s}_{L,n}\right)=\mathcal{D}(\boldsymbol{s}_{L,n};\boldsymbol{\beta}_n),\quad n=1,...,N.
\end{equation}
Then the ML problem (\ref{eq:ML}) is approximately solved by maximizing the lower bound,
\begin{equation}
\begin{aligned}\label{eq:approx ml}
\max_{\Theta\in\mathcal{C}, \boldsymbol{B}\geq 0} \frac{1}{N}\sum_{n=1}^N\mathbb{E}_{\boldsymbol{s}_{L,n}\sim q_n} \left[\log\frac{ p(\boldsymbol{y}_n,\boldsymbol{s}_{L,n} ;\Theta)}{ q(\boldsymbol{s}_{L,n};\boldsymbol{\beta}_n)}\right],
\end{aligned}
\end{equation}
where $\boldsymbol{B}$ collects {\revcolor the $\boldsymbol{\beta}_n$'s as its columns}.
We note that the objective in (\ref{eq:approx ml}) 
is essentially the same as the one in {\revcolor our prior work} \cite{wu2021probabilistic} which admits a closed-form expression.

We tackle problem \eqref{eq:approx ml} via alternating maximization \wrt $\boldsymbol{B}$ and all parameters in $\Theta$.
To be specific, we update $\boldsymbol{A}_1$ and $\boldsymbol{S}_l$'s via accelerated proximal gradient descent \cite{beck2009fast}; $\sigma^2$ can be updated in closed-form expression; for the variational parameter $\boldsymbol{B}$, we follow the same way as that in \cite{wu2021probabilistic}.
We omit the details due to the limited space.

For the initialization of the parameters, we use one-layer matrix factorization methods to initialize the top layer and the bottom layer.
To be specific, we use VCA \cite{nascimento2005vertex} and FCLS \cite{heinz2001fully} to retrieve $\boldsymbol{A}_1$ and $\boldsymbol{S}_L$ from $\boldsymbol{Y}$ by the following factorization
\begin{equation}
    \boldsymbol{Y}=\boldsymbol{A}_L\boldsymbol{S}_L,\quad \boldsymbol{Y}=\boldsymbol{A}_1\boldsymbol{\hat S},
\end{equation}
where $\boldsymbol{\hat S}\in\mathbb{R}_+^{N_1\times N}$ and $\boldsymbol{\hat S}^\top\boldsymbol{1}=\boldsymbol{1}$.
Then, we initialize $\boldsymbol{B}$ by $\boldsymbol{S}_L$ and initialize $\boldsymbol{S}_1,\dots,\boldsymbol{S}_{L-1}$ by randomly sampling the columns from the uniform Dirichlet distribution.

\begin{figure}[htb!]
	\centering
	\subfigure[The sub-image]{
		\begin{minipage}[]{0.15\textwidth}
			\includegraphics[width=1\textwidth]{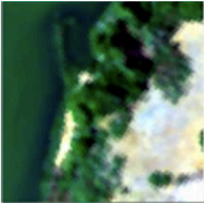}
		\end{minipage}
	}
    	\subfigure[Reference EMs]{
    		\begin{minipage}[]{0.23\textwidth}
   		 	\includegraphics[width=1\textwidth]{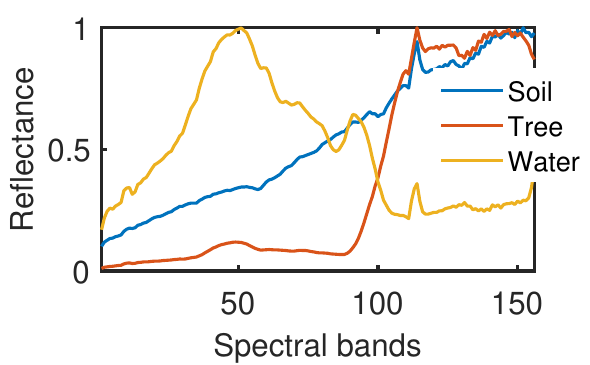}
    		\end{minipage}
    	}
\caption{ The sub-image and three materials in the scene.}
\label{fig:ref}
\end{figure}

\begin{figure*}[htb]
    \centering    \includegraphics[scale=0.25]{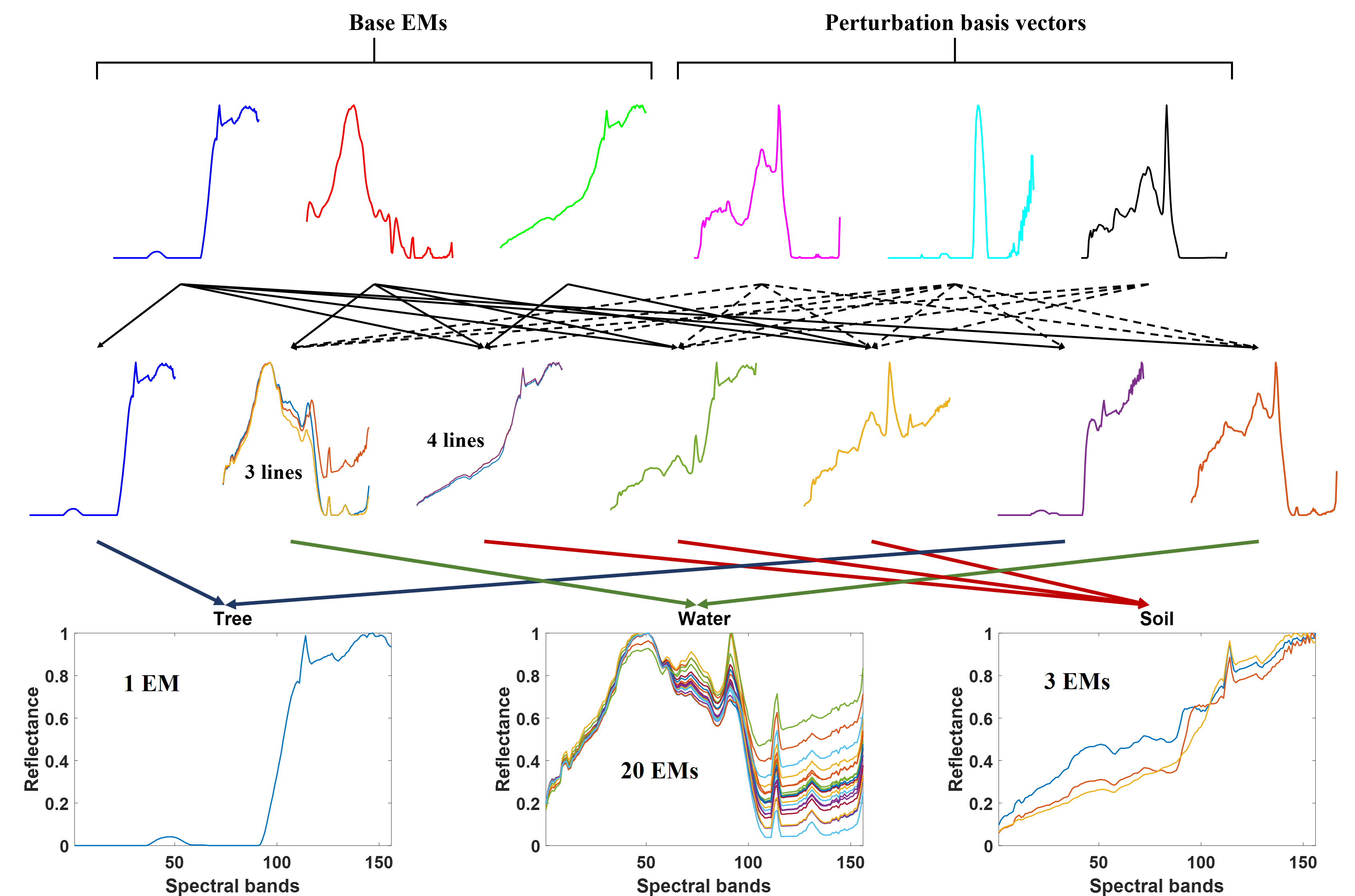}
    \caption{From top to down: columns of $\boldsymbol{A}_1$, $\boldsymbol{A}_2$, and $\boldsymbol{A}_3$.
    Arrows indicate convex combinations.}
    \label{fig:A123}
\end{figure*}
\begin{figure}[htb!]
    \centering 
    \includegraphics[scale=0.15]{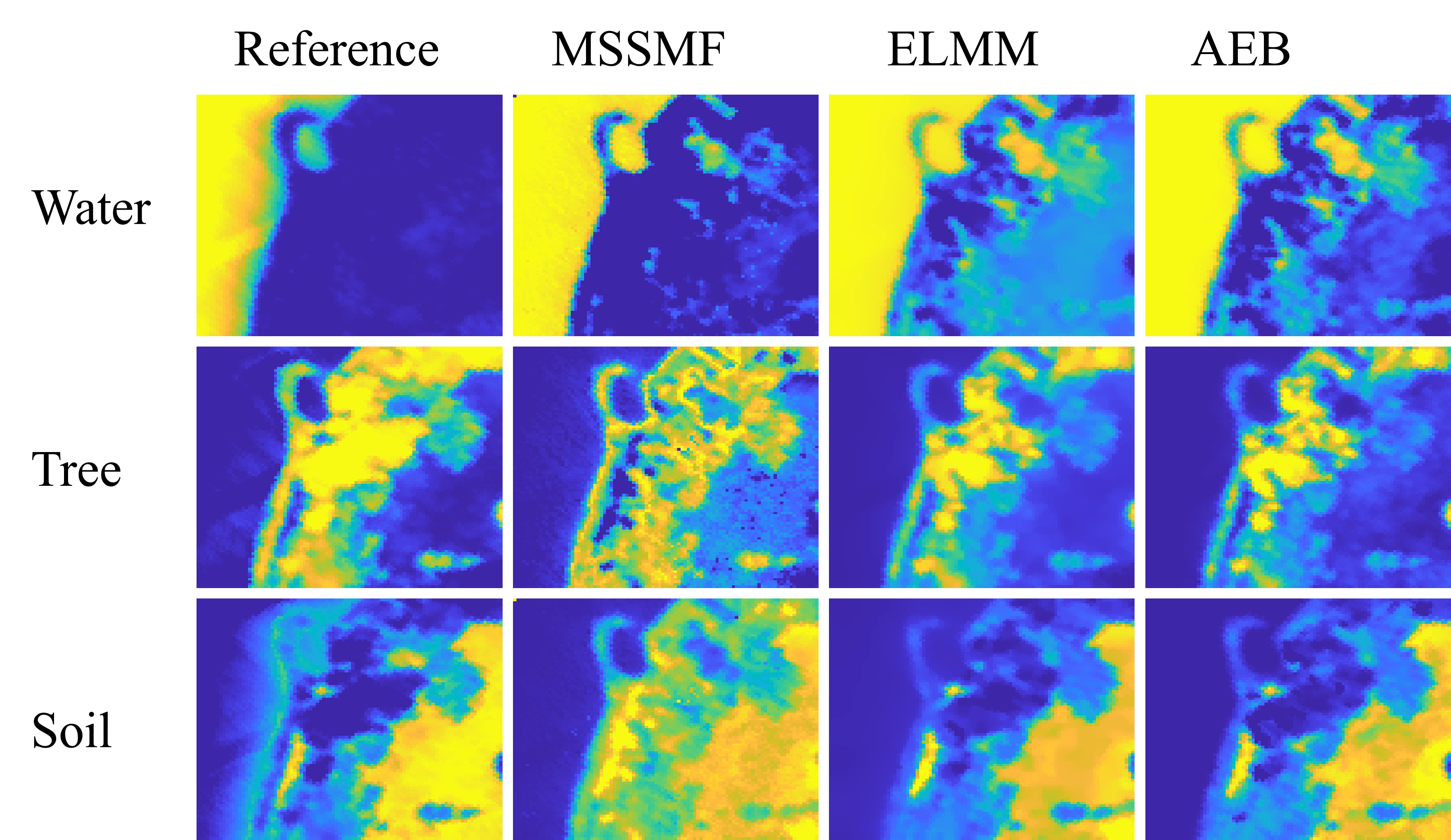}
    \caption{Estimated abundance maps of the \textit{Samson} image.}
    \label{fig:abd}
\end{figure}

\begin{figure*}[htb]
	\centering
	\subfigure[\scalebox{0.95}{MSE vs. SNR}]{
		\begin{minipage}[]{0.26\textwidth}
			\includegraphics[width=1\textwidth]{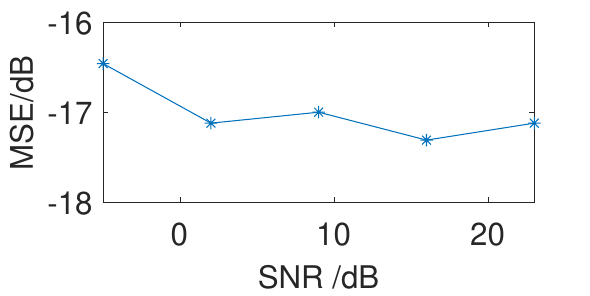}
		\end{minipage}
	}
     	\subfigure[\scalebox{0.95}{Endmembers extracted by MSSMF; SNR$=20$dB.}]{
		\begin{minipage}[]{0.26\textwidth}
			\includegraphics[width=1\textwidth]{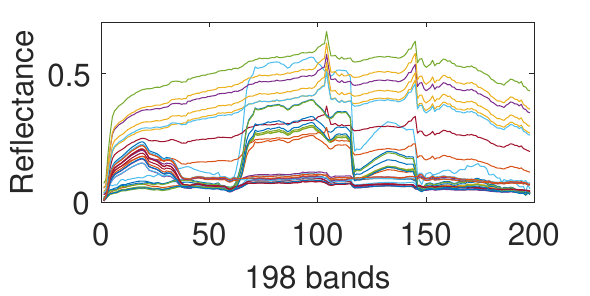}
		\end{minipage}
	}
    	\subfigure[\scalebox{0.95}{Ground truth.}]{
    		\begin{minipage}[]{0.26\textwidth}
   		 	\includegraphics[width=1\textwidth]{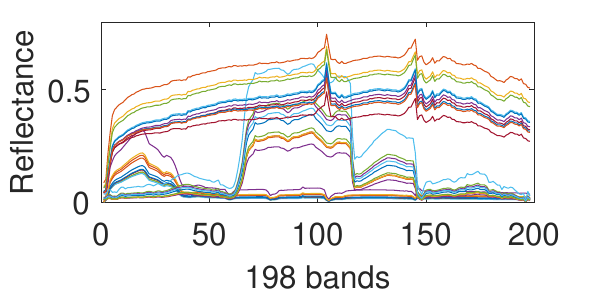}
    		\end{minipage}
    	}
\caption{Results of synthetic data experiments.}
\label{fig:simu}
\end{figure*}

\section{Experiment}
\subsection{Real hyperspectral image experiment}
We validate the proposed MSSMF for HU with EV through experiments on the Samson hyperspectral image.
We take a sub-image with size $95\times 95$.
The {\revcolor RGB plot of the sub-image and the reference EMs of} the three presented materials are shown in Fig. \ref{fig:ref}.

We choose two HU methods that also consider EV as benchmarks: extended linear mixture model (ELMM) \cite{drumetz2016blind} and automated endmember bundles (AEB) \cite{somers2012automated}.
We use the source code of ELMM and implement AEB by ourselves. 
The model dimension of MSSMF, \textit{i.e.} $(M, N_1,...,N_L)$, is set as $(156, 6, 12, 24, 95\times 95)$.
We stop MSSMF after $100$ iterations.

The estimated abundance maps are shown in Fig. \ref{fig:abd}.
We see that MSSMF arguably gives more detailed distributions of the materials that are consistent with visual observation on the {\revcolor sub-image}.

It is interesting to see what EMs MSSMF retrieved.
The $24$ EMs unmixed by MSSMF, {\revcolor the} columns of $\boldsymbol{A}_2$, and the base EMs and the perturbation basis are plotted in Fig. \ref{fig:A123}.
The figure {\revcolor shows} how the final EMs are hierarchically generated from the core
basis matrix.
We see that the core
basis matrix indeed captures {\revcolor the} three base EMs in the shape of spectral signatures of water, soil, and tree.
Also, there are three perturbation basis vectors accounting for variability.
Through hierarchical convex combinations, 24 EMs are obtained where water has the most {\revcolor varied} EMs while tree only has one.

\subsection{Synthetic data experiments}
We perform synthetic data experiments to test MSSMF.
We choose three endmembers with 198 spectral bands which are the same as those used in Fig. \ref{fig:singular_val}. We then generate $200$ variants for each EM to form an EM collection following the same EV generation procedure {\revcolor in} \cite{borsoi2021spectral}.
We randomly pick out $10$ for each of {\revcolor the} three EMs to form the ground truth $\boldsymbol{A}^\star$.

The data matrix is obtained by {\revcolor the LMM}, $\boldsymbol{Y}=\boldsymbol{A}^\star\boldsymbol{Z}^\star+\boldsymbol{V}$, where the columns of the abundance matrix are generated from {\revcolor the} uniform Dirichlet distribution; the noise variance $\sigma^2$ is set such that the given signal-to-noise ratio (SNR) is satisfied.
{\revcolor The} SNR is defined as $\|\boldsymbol{AZ}\|_F^2/(\sigma^2MN)$.


We vary the noise level to test MSSMF. 
The model dimension of MSSMF is set as $(198, 6, 18, 30, 2500)$. 
We perform 50 Monte Carlo runs.
We choose the mean square error (MSE) as the metric:
\begin{equation}
    \text{MSE}(\boldsymbol{A}^\prime, \boldsymbol{ A}^\star):=\|\boldsymbol{A}^\prime\boldsymbol{\Pi}-\boldsymbol{A}^\star\|_F^2/\left(K\|\boldsymbol{A}^\star\|_F^2\right),
\end{equation}
where $\boldsymbol{A}^\prime$ refers to the estimation returned by the algorithm; $\boldsymbol{\Pi}$ is a permutation matrix that aligns the columns. 
The MSE performance is shown in Fig. \ref{fig:simu}(a) and it shows that MSSMF gives reasonable MSE performance.
Fig. \ref{fig:simu}(b) showcases the extracted EMs of a random realization.
The result is visually reasonable compared to the ground truth Fig. \ref{fig:simu}(c).

\section{Conclusion}
We {\revcolor studied} a multilayer simplex-structured matrix factorization model for HU with EV in this paper. 
{\revcolor Experimental results suggest that multilayer structured factorization may be used to account for EV, which is a challenging aspect in HU.}
As future work, it would be interesting to study the effects of the number and {\revcolor sizes of the} layers and the identifiability of the model.

\bibliographystyle{ieeetr}
\bibliography{refs-JB, refs-YN}

\end{document}